\documentclass[]{spie}  %>>> use for US letter paper
\usepackage{amsmath,amsfonts,amssymb}
\usepackage{graphicx}
\usepackage{url}
\usepackage{float}
\newcommand{\tabincell}[2]{\begin{tabular}{@{}#1@{}}#2\end{tabular}}
\usepackage[colorlinks=true, allcolors=blue]{hyperref}

\title{Design of microstrip-line coupled kinetic inductance detectors for near infrared astronomy}

\author[ab]{Shiling Yu}
\author[c*]{Shibo Shu}
\author[a$\dagger$]{Ran Duan}
\author[d]{Lihui Yang}
\author[a]{Di Li}

\affil[a]{National Astronomical Observatories, Chinese Academy of Sciences, Beijing 100101, China}
\affil[b]{School of Astronomy and Space Science, University of Chinese Academy of Sciences, Beijing 100049, China}
\affil[c]{Institute of High Energy Physics, Chinese Academy of Sciences, Beijing 100049, China}
\affil[d]{Zhejiang Lab, Hangzhou, Zhejiang 311121, China}
\affil[*]{shusb@ihep.ac.cn}
\affil[$\dagger$]{Send correspondence to duanran@nao.cas.cn}

\begin{document} 
\maketitle

\begin{abstract}

Kinetic inductance detectors (KID) have great potential in astronomical observation, such as searching for exoplanets, because of their low noise, fast response and photon counting characteristics. In this paper, we present the design process and simulation results of a microstrip line coupled KIDs array for near-infrared astronomical observation. Compared with coplanar waveguide (CPW) feedlines, microstrip feedlines do not require air bridges, which simplify fabrication process. In the design part, we mainly focus on the impedance transforming networks, the KID structure, and the frequency crosstalk simulations. The test array has a total of 104 resonators with 8 rows and 13 columns, which ranges from 4.899~GHz to 6.194~GHz. The pitch size is about 200~$\mu$m and the frequency crosstalk is less than 50~kHz in simulation.

\end{abstract}


\section{Introduction}
\label{sec:intro}  % \label{} allows reference to this section

In the "Pathways to Discovery in Astronomy and Astrophysics for the 2020s", the main scientific research topics can be summarized in three directions: exploring the livable world, the dynamic universe, and the galaxy evolution~\cite{NAP26141} . Among them, the search for exoplanets, the characteristics and structure of dark matter halo, and the chemical evolution of galaxies all need and rely on the improvement of infrared detector technology. For example, direct imaging will provide the profile of the exoplanet atmosphere, and may also provide the present evidence of gases that could indicate life~\cite{NASAexoplanets} . So far a total of 5035 exoplanets have been found, but only 59 have been found by direct imaging~\cite{NASAexoplanets} .

Kinetic inductance detector emerges as the times require. For the high contrast direct imaging of exoplanets, KID will actively suppress speckles because of its high near-infrared sensitivity, lack of reading noise and fast reading characteristics~\cite{benmazin19} . In addition, KID can count photons and has the intrinsic spectral resolution. At present, the astronomical instruments using KID in the near-infrared band include ARCONS~\cite{Mazin2013} , DARKNESS~\cite{Meeker2018} , and MEC~\cite{2019PhDW, Walter2020} . All these instruments use CPW feedlines to realize intrinsic frequency domain multiplexing readout~\cite{Mazin2013, Meeker2018, Walter2020} . The fabrication of CPW coupled KID arrays requires air bridges to suppress other CPW modes~\cite{Szypryt17} . Using microstrip feedlines can simplify the fabrication process. Therefore, we demonstrate the design and simulation results of a microstrip-line coupled KIDs test array for near-infrared detection.

\subsection{Basic principle of KID}

When the photon energy absorbed by the superconducting material is higher than the gap energy, Cooper pairs will be broken, which changes the inductance impedance and shifts the resonance frequency to lower frequency. The change of the resonance frequency depends on the number of Cooper pairs destroyed by the incident photons, so it is directly proportional to the energy deposited in the superconductor. By reading out the change values of resonators, the photon position, arrival time, number and energy information can be obtained~\cite{2003Nature, benmazin19} .  

The gap energy is calculated as $2\Delta =3.52 k_B T_c$, where $k_B$ is the Boltzmann constant, $T_c$ is the transition temperature, and $\Delta$ is the gap energy. Taking aluminum film as an example, $T_c$ is 1.18~K, and 2$\Delta$ is about  3.582 $\times 10^{-4}$~eV. Therefore, photons can be detected as long as their wavelength is shorter than 3.46~mm. Near infrared photons with wavelength range of 780~nm - 3000~nm is 0.41~eV - 1.59~eV, much larger than the gap energy of usual superconductors.

\section{Design and simulation}
\label{sec:design}

The most critical structure in the design of a KIDs array is the superconducting resonator. The effect of resonator pixel size, pitch size and arrangement on frequency crosstalk can be simulated by Sonnet~\cite{sonnet} . 

The design parameters are set as a 18~nm upper aluminum film with $L_s$ = 2~pH/sq, a 150 $\mu$m-thick high resistance silicon with a dielectric constant of 11.7, and a 200~nm aluminum film as ground. In the future, we will switch to low reflectivity superconducting materials. The simulation setup detail are shown in Appendix~\ref{sec:Preparation} and Appendix~\ref{sec:MRL}. The design includes four main parts, the impedance transforming networks, the KIDs, the frequency crosstalk simulation, and the picth and arrangement.

\subsection{Impedance transforming networks}

\begin{figure} [ht]
\centering
\begin{minipage}[b]{0.48\textwidth}
\centering
\includegraphics[height=3.5cm,width=6cm]{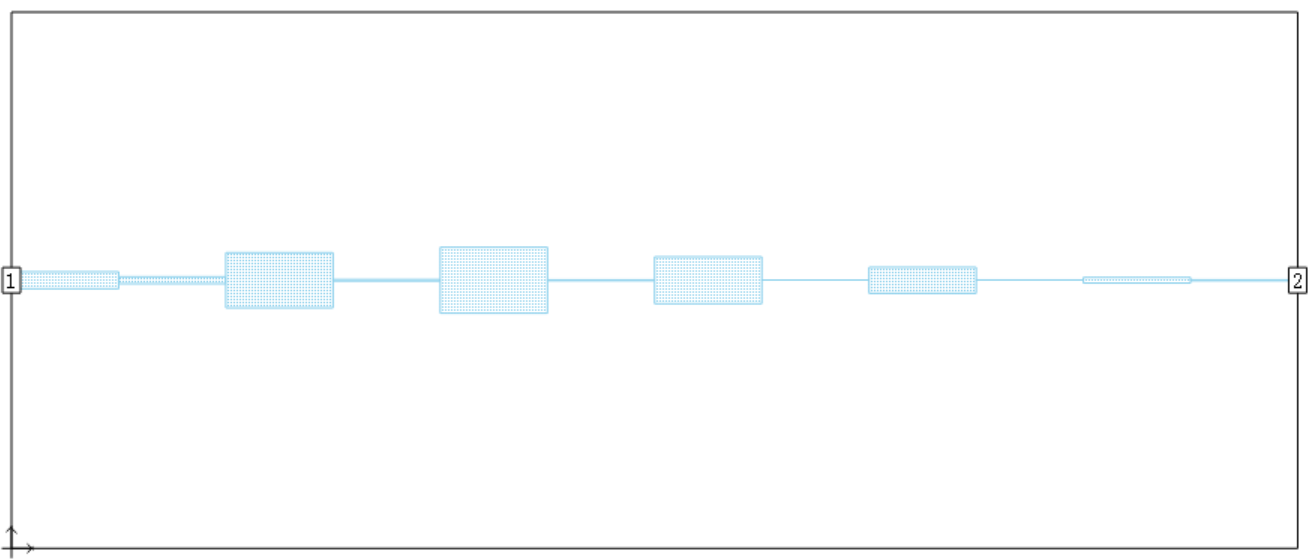}
\caption{Impedance transforming networks}
\label{ImpedanceTransformationPart}
\end{minipage}
\begin{minipage}[b]{0.48\textwidth}
\centering
\includegraphics[height=4.2cm,width=6cm]{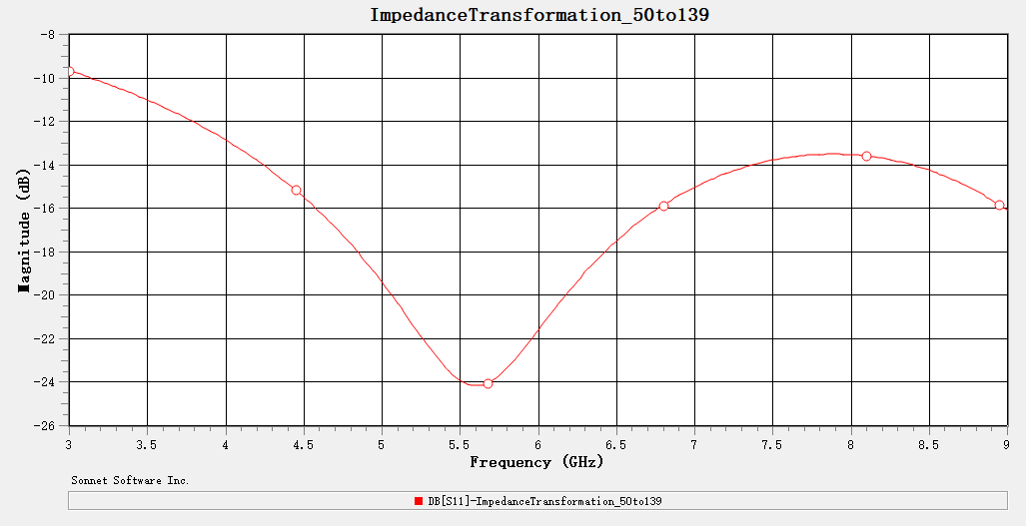}
\caption{Simulation results $S_{11}$}
\label{ImpedanceTransformationS11}
\end{minipage}
\end{figure}

The width of the microstrip determines the characteristic impedance of the feedline. To make the pixels compact, the microstrip width is decreased to 6 $\mu$m, and the corresponding characteristic impedance value is 140~$\Omega$. To match the 50~$\Omega$ cables, an impedance transformation network is used~\cite{1966Matthaei} and the calculation process is shown in Appendix~\ref{sec:ITN}. The network design has 10 sections, which have the same length but different impedance, shown in Fig.~\ref{ImpedanceTransformationPart}. The simulation result in Fig.~\ref{ImpedanceTransformationS11} shows that $S_{11}$ is less than -10~dB in the range of 4~GHz - 8~GHz. 

\subsection{Superconducting resonators}
\label{sec:SR}

\begin{figure} [ht]
\centering
\begin{minipage}[b]{0.48\textwidth}
\centering
\includegraphics[width=6cm]{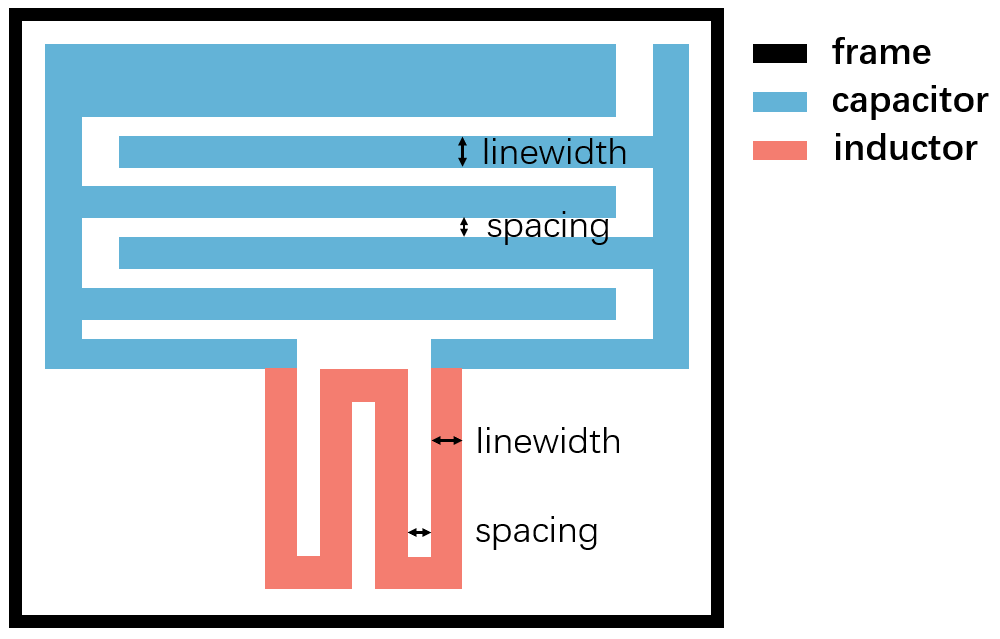}
\caption{Design Diagram}
\label{fig:sketch}
\end{minipage}
\begin{minipage}[b]{0.48\textwidth}
\centering
\includegraphics[width=8.2cm]{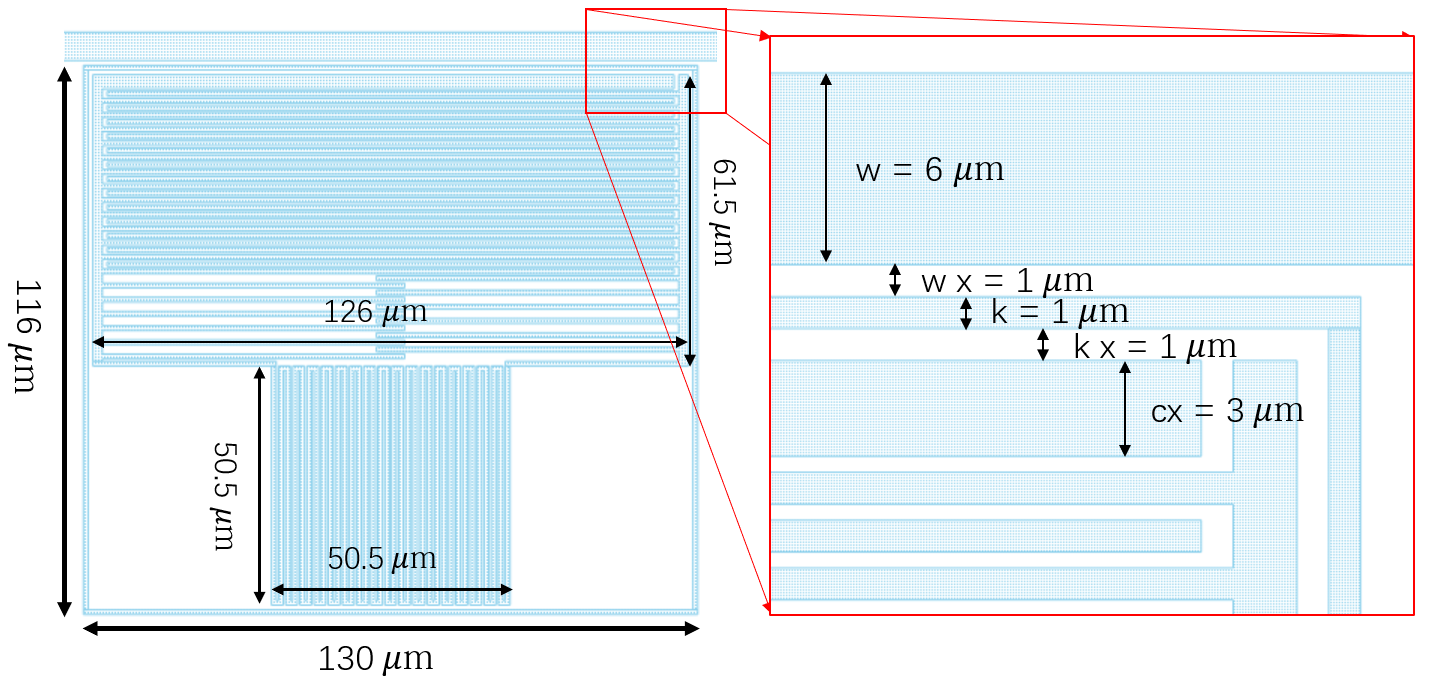}
\caption{Description of Resonator Size}
\label{fig:design}
\end{minipage}
\end{figure}

The resonators are coupled capacitively to the microstrip line. The design of a lumped-element resonator is composed of a meandered inductor and an interdigitated capacitor~\cite{Jonas12} , both of them with 1 $\mu$m linewidth and 0.5 $\mu$m spacing, as shown in Fig.~\ref{fig:sketch}. In addition, a rectangular frame is added to reduce the frequency crosstalk ~\cite{shu18} . The inductor is the main sensitive part of the photon event, with the size of 50.5 $\mu$m $\times$ 50.5 $\mu$m. The interdigitated capacitor contains 19 pairs of fingers to adjust different resonant frequencies, with the size of 126 $\mu$m $\times$ 61.5 $\mu$m. The finger length varies from 64 $\mu$m to 121 $\mu$m, and the adjustable frequency range is about from 4~GHz to 10~GHz. The overall size is 130 $\mu$m $\times$ 116 $\mu$m, as shown in Fig.~\ref{fig:design}.

The detail parameters are optimized to make the resonator have the appropriate quality factor. The quality factor is defined as the resonant frequency multiplied by the ratio of the average stored energy to the energy loss. The relationship among the total quality factor $Q$, the coupling quality factor $Q_c$ and the internal quality factor $Q_i$ is as follows:

\begin{equation}
\label{eq:Q}
\begin{aligned}
\frac{1}{Q}=\frac{1}{Q_i}+\frac{1}{Q_c}
\end{aligned}
\end{equation}

$Q_c$ represents the coupling between the resonator and the feedline. By testing the effect of the five variables shown in Fig.~\ref{fig:design} on the quality factor $Q_c$, the design values are determined as ($\mu$m): $w = 6, wx = 1, k = 1, kx = 1, cx = 3$. The resonance frequency is between 4.899~GHz and 6.194~GHz, and the corresponding quality factor $Q_c$ is between 51.0~k and 25.3~k.

\subsection{Frequency crosstalk simulation}

In order to reduce the interaction between pixels, it is necessary to simulate the frequency crosstalk between resonators. Here, we consider three configurations, which have the greatest impact on the reference resonator, the horizontal direction, the vertical direction and the resonators placed on two sides of the feedline~\cite{shu18} . The frequency crosstalk is defined as the frequency shifts between the resonance frequencies simulated in a single resonator setup and a two-resonator setup. 

\subsubsection{Crosstalk between resonators in the horizontal direction} 

\begin{figure} [ht]
\centering
\begin{minipage}[b]{0.48\textwidth}
\centering
\includegraphics[width=8cm]{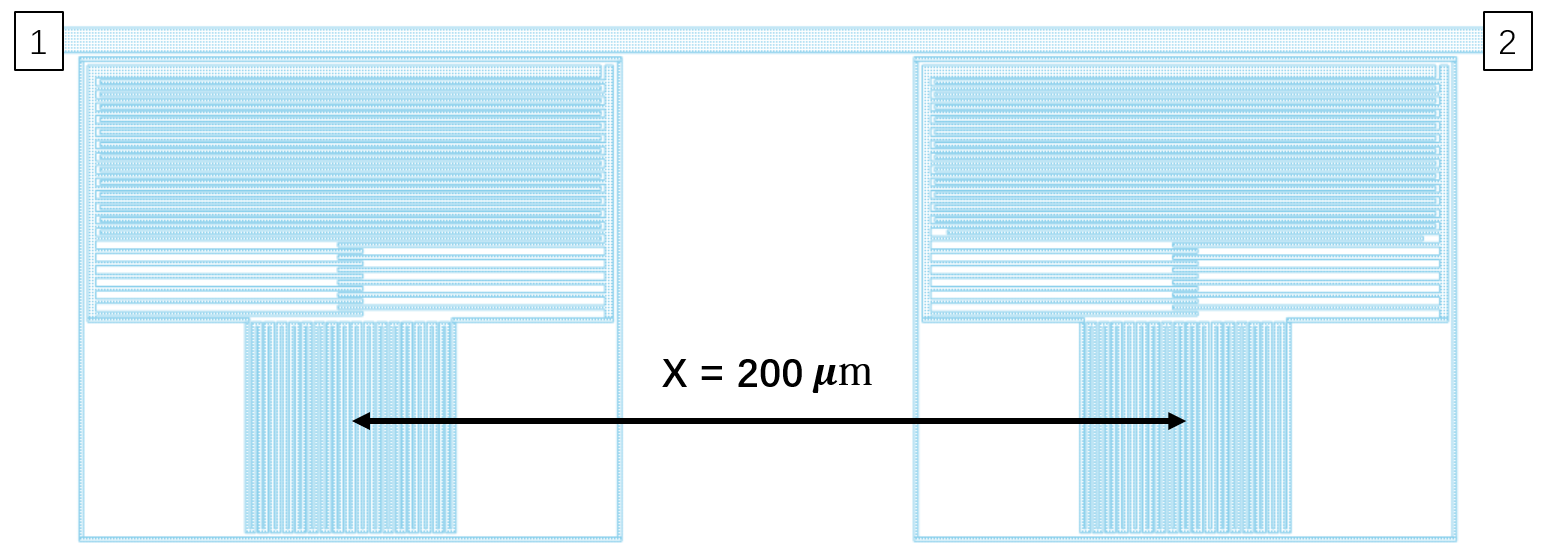}
\caption{Simulation setup in horizontal direction}
\label{xpicture}
\end{minipage}
\begin{minipage}[b]{0.48\textwidth}
\centering
\includegraphics[width=8.2cm]{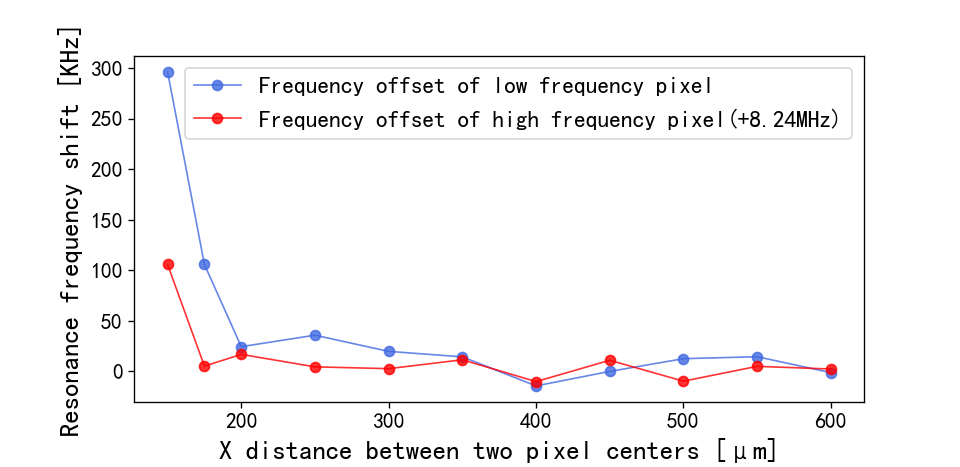}
\caption{The crosstalk with different X distance}
\label{xdirection}
\end{minipage}
\end{figure}

In Fig.~\ref{xpicture}, two resonators with frequency difference of 8~MHz are placed at a horizontal distance of 200 $\mu$m. Fig.~\ref{xdirection} shows that when $X$ is larger than 200 $\mu$m, the crosstalk is almost constant and less than 40~kHz.

\subsubsection{Crosstalk between resonators in the vertical direction}

\begin{figure} [ht]
\centering
\begin{minipage}[b]{0.48\textwidth}
\centering
\includegraphics[width=4.5cm]{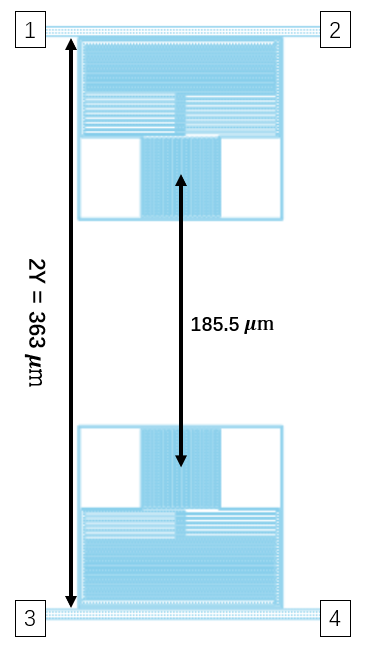}
\caption{Simulation setup in the vertical direction}
\label{ypicture}
\end{minipage}
\begin{minipage}[b]{0.48\textwidth}
\centering
\includegraphics[width=8.2cm]{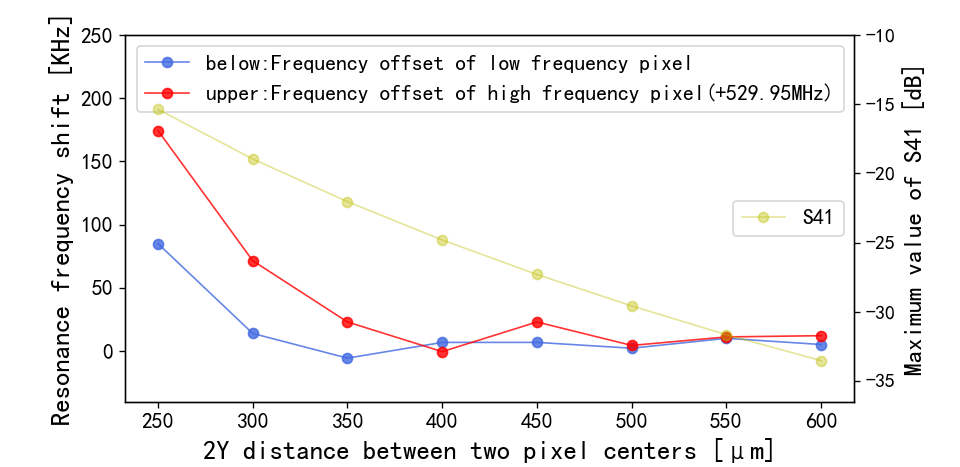}
\caption{The simulated crosstalk by varying $2Y$.}
\label{ydirection}
\end{minipage}
\end{figure}

In Fig.~\ref{ypicture}, two resonators with frequency difference of 530~MHz are placed at a vertical distance of 363 $\mu$m (the distance between inductance parts is 185.5 $\mu$m). Fig.~\ref{ydirection} shows the simulated crosstalk by varying $2Y$. When $2Y>350$~$\mu$m, the crosstalk is almost constant around 35~kHz. The maximum value of $S_{41}$ curve obtained by simulation shows the maximum crosstalk between the two feedlines, which is smaller than -22~dB when $2Y>350$~$\mu$m. 

\subsubsection{Crosstalk between resonators on two sides of the feedline}

\begin{figure} [ht]
\centering
\begin{minipage}[b]{0.48\textwidth}
\centering
\includegraphics[width=4cm]{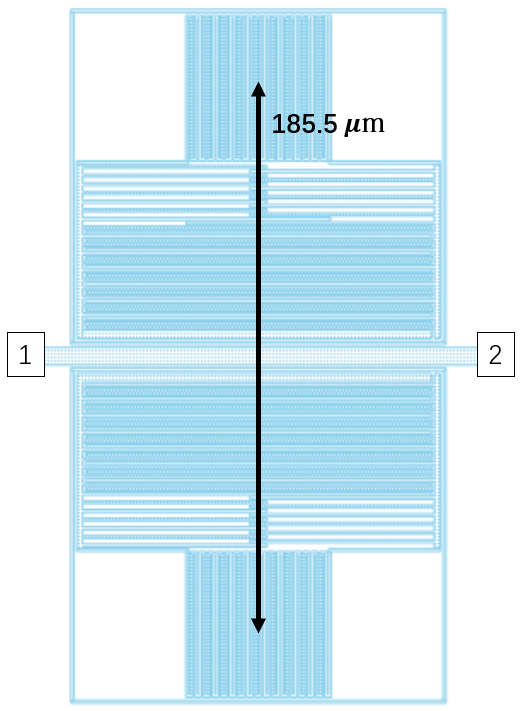}
\caption{Simulation setup on two sides of the feedline}
\label{oppospicture}
\end{minipage}
\begin{minipage}[b]{0.48\textwidth}
\centering
\includegraphics[width=8.2cm]{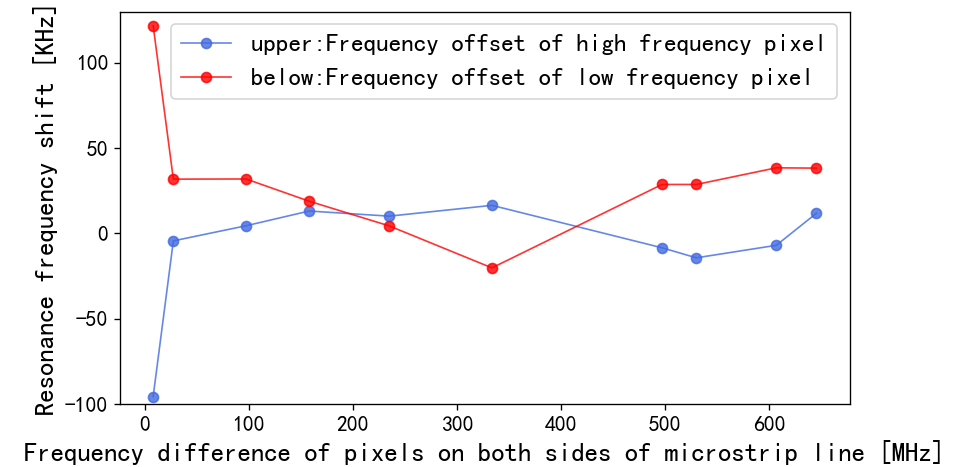}
\caption{The crosstalk with different resonance frequency}
\label{oppps}
\end{minipage}
\end{figure}

In Fig.~\ref{oppospicture}, two resonators with different resonance frequencies are placed on two sides of the feedline, and the distance between the inductance parts is 185.5 $\mu$m. Here we consider the crosstalk between resonators on two sides of the feedline by varying the resonance frequency difference between these two resonators. Fig.~\ref{oppps} shows that the crosstalk is almost constant, around 40~kHz, when the frequency difference is larger than 27.6~MHz. The fluctuation of the simulated crosstalk may come from the numerical error when the cell size is much smaller than the wavelength in Sonnet.

\subsection{Pitch size and pixel arrangement}

\begin{figure} [ht]
   \begin{center}
   \begin{tabular}{c} 
   \includegraphics[height=7cm]{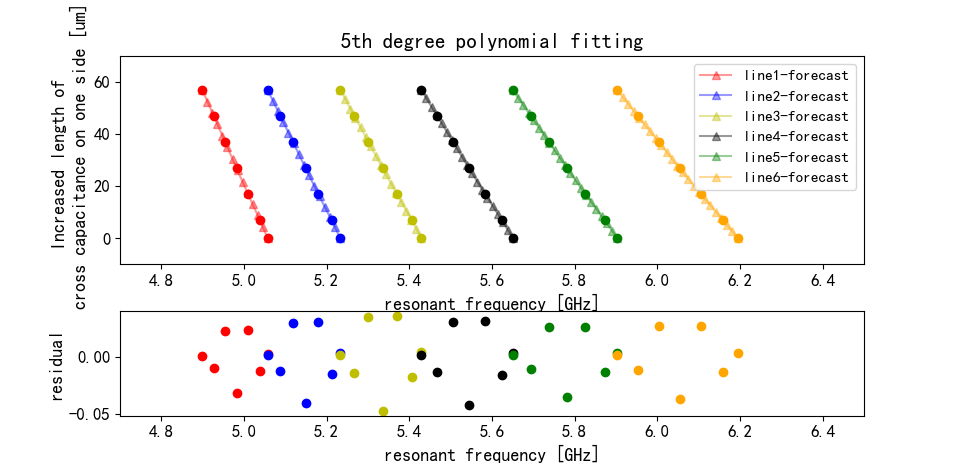}
	\end{tabular}
	\end{center}
   \caption[example] 
   { \label{fig:sixline5} 
Polynomial fitting results between the length of capacitor fingers and the resonance frequency~\cite{Shu2021} .}
   \end{figure}

\begin{figure} [ht]
   \begin{center}
   \begin{tabular}{c} 
   \includegraphics[height=7cm]{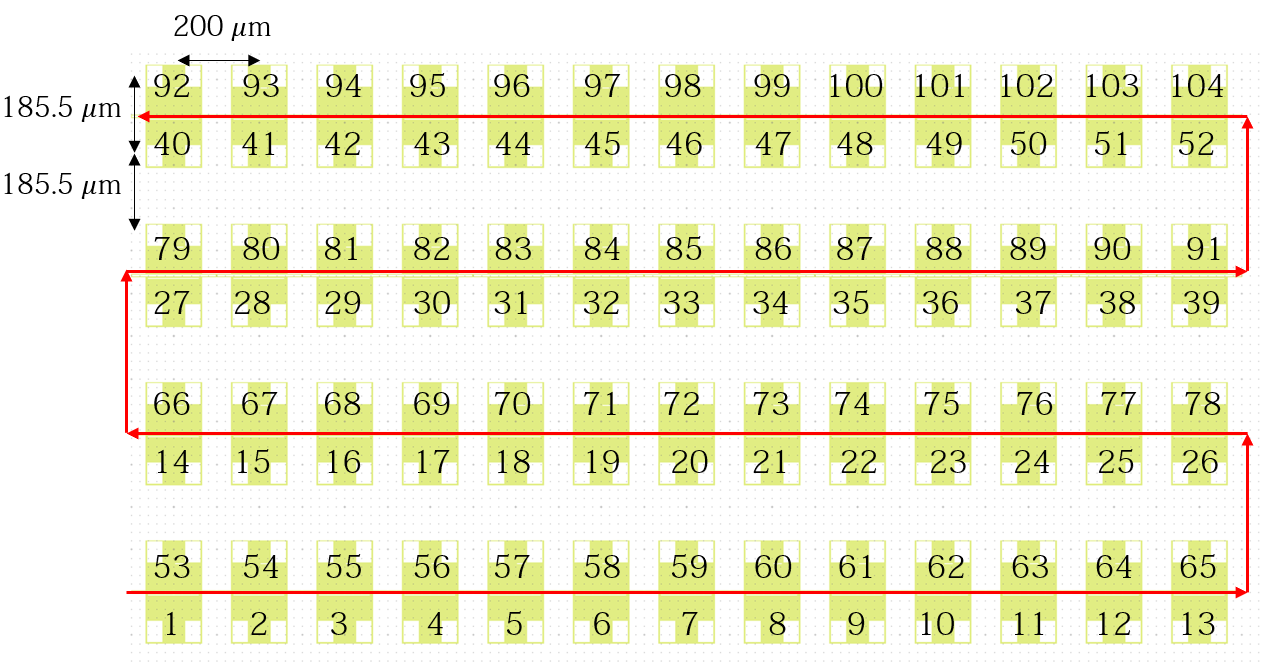}
	\end{tabular}
	\end{center}
   \caption[example] 
   { \label{fig:arrangement} 
The pitch sizes and pixel arrangement of 104 pixels. The index shows the frequency numbering from 4.899~GHz to 6.194~GHz.}
   \end{figure}

A test array with 104 pixels is designed for testing. To minimize the crosstalk, the horizontal pitch size $X$ is set to be 200 $\mu$m. To have a uniform pitch size in the vertical direction, $2Y$ is set to be 363 $\mu$m and the vertical pitch size is 185.5~$\mu$m. To have $Q_c$ around 30~k, the resonance frequency range of 4.899~GHz - 6.194~GHz is used, with a resonance frequency spacing of 12.57~MHz. The frequency difference between the resonators on two sides of the feedline is set to be 654~MHz, the half readout frequency range in our test array. The capacitor finger lengths are interpolated from the results in Fig.~\ref{fig:sixline5}. The final test array has 104 pixels in 8 rows and 13 columns, as shown in Fig.~\ref{fig:arrangement}.

\section{CONCLUSION}

We present a microstrip line coupled KID array design for near-infrared astronomy. An impedance transformation network is used to match the 140 $\Omega$ feedline impedance to the 50 $\Omega$ cable. The size of a single resonator is 130 $\mu$m $\times$ 116 $\mu$m, with a square inductor of 50.5 $\mu$m $\times$ 50.5 $\mu$m. The horizontal and vertical pitch sizes are set to be 200 $\mu$m and 185.5 $\mu$m, respectively. The frequency crosstalks are simulated in three configurations, and the simulation results are less than 50~kHz. A test array of 104 resonators in 8 rows and 13 columns ranges from 4.899~GHz to 6.194~GHz with the frequency interval of 12.57~MHz, and the $Q_c$ value range is 25.3~k - 51.0~k. And the measurement of photon response and noise equivalent power is in progress. 

% Note: If compiling with LaTeX+dvipdf, please ensure images generated from
% other software packages have their bounding boxes set correctly.
%>>>> use \label inside caption to get Fig. number with \ref{}

\appendix    %>>>> this command starts appendixes

\section{Preparation}
\label{sec:Preparation}

Due to the limitation of fabrication, aluminum thin film is selected as the superconducting material. Considering the high reflectivity of aluminum, other superconducting materials with high absorptivity will be used in the future. 

The microstrip transmission line is composed of a $w$ width conductive strip on the surface of a dielectric plate with a thickness of $h$, and a layer of conductive grounding is under the dielectric plate~\cite{Jonas12} . The fabrication parameters are designed as: 18~nm upper aluminum film with $L_s$ = 2~pH/sq; 150 $\mu$m ($h$ = 150 $\mu$m) high resistance silicon dielectric with a dielectric constant of 11.7; 200~nm bottom aluminum film grounding.

Before the formal design simulation, the appropriate basic unit size needs to be set. Too large cell size will lead to inaccurate simulation results, and too small cell size will take more time. Therefore, it is necessary to test the required memory and simulation results of different cell sizes.

As shown in Tab.~\ref{tab:preparation}, taking the resonance frequency obtained when cell size is 0.05 $\mu$m as the reference value, two resonators with different resonant frequencies ( about 4GHz and 8GHz ) were tested. Considering the time cost and accuracy, the cell size is finally set to 0.1 $\mu$m and the mesh of Sonnet choose "Fine/Edge Meshing".

\begin{table}[ht]
\caption{The test results with different basic cell sizes and different mesh of Sonnet, include the estimation of memory (determining the simulation time) and frequency differences (high-frequency resonator and low-frequency resonator are tested respectively).} 
\label{tab:preparation}
\begin{center}       
\begin{tabular}{|l|l|l|l|l|l|l|}
\hline
\rule[-1ex]{0pt}{3.5ex}  cell size [$\mu$m] & 0.05 & 0.1 & 0.1 & 0.1 & 0.25 & 0.5  \\
\hline
\rule[-1ex]{0pt}{3.5ex}  mesh & \tabincell{l}{Fine/Edge\\ Meshing} & \tabincell{l}{Fine/Edge\\ Meshing} & \tabincell{l}{Coarse/Edge\\ Meshing} & \tabincell{l}{Coarse/No\\Edge Meshing} & \tabincell{l}{Fine/Edge\\ Meshing} & \tabincell{l}{Fine/Edge\\ Meshing} \\
\hline
\rule[-1ex]{0pt}{3.5ex}  \tabincell{l}{estimated\\memory} & 11906MB & 3213MB & 578MB & 362MB & 534MB & 120MB  \\
\hline
\rule[-1ex]{0pt}{3.5ex}  $F_{min}$ [GHz] & 3.97849 & 4.00111 & 3.99979 & 4.19353 & 4.06933 & 4.18769  \\
\hline
\rule[-1ex]{0pt}{3.5ex}  $\Delta$ [MHz] & 0 & 22.62 & 21.3 & 215.04 & 90.84 & 209.2  \\
\hline 
\rule[-1ex]{0pt}{3.5ex}  $F_{max}$ [GHz] & 7.95573 & 7.9995 & 8.19636 &   & 8.15001 & 8.34143  \\
\hline
\rule[-1ex]{0pt}{3.5ex}  $\Delta$ [MHz] & 0 & 43.44 & 240.63 &   & 194.28 & 385.7  \\
\hline 
\end{tabular}
\end{center}
\end{table}

\section{Microstrip Feedline}
\label{sec:MRL}

A 50~$\Omega$ microstrip feedline with 130 $\mu$m width seriously hinders the compact arrangement of a KIDs array. It is necessary to make a narrow feedline design. An impedance transformation network is needed for this high impedance feedline.
From Tab.~\ref{tab:MTTLwidth}, the width is finally set to 6 $\mu$m.

\begin{table}[ht]
\caption{The test results of feedline width and impedance, and the length of feedline is taken as 500 $\mu$m during test.} 
\label{tab:MTTLwidth}
\begin{center}       
\begin{tabular}{|l|l|}
\hline
\rule[-1ex]{0pt}{3.5ex}  Microstrip linewidth [$\mu$m] & Characteristic impedance[$\Omega$]  \\
\hline
\rule[-1ex]{0pt}{3.5ex}  6 & 136 \\
\hline
\rule[-1ex]{0pt}{3.5ex}  8 & 124 \\
\hline
\rule[-1ex]{0pt}{3.5ex}  10 & 116 \\
\hline
\rule[-1ex]{0pt}{3.5ex}  126 & 59 \\
\hline
\end{tabular}
\end{center}
\end{table}

\begin{figure} [ht]
  \begin{center}
  \begin{tabular}{c} 
  \includegraphics[height=7cm]{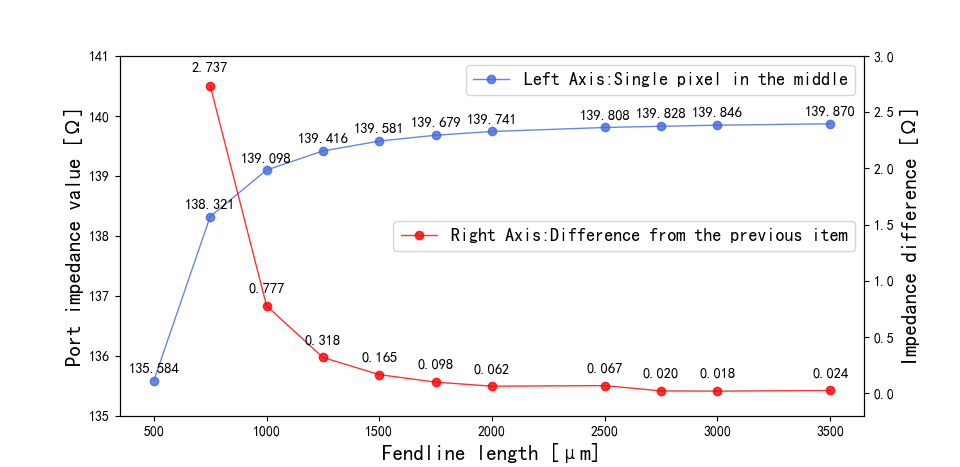}
	\end{tabular}
	\end{center}
  \caption[example] 
  { \label{fig:linelength} 
The impedance value gradually tends to a stable value with the increase of the length of the feedline, and the difference with the previous term tends to zero. During the test, the resonator is placed in the middle regardless of the length of the feedline. }
  \end{figure}

If the length of the feedline is too short, the impedance value will be inaccurate. When the width is 6 $\mu$m, the impedance value changes with the different lengths of feedline, as shown in Fig.~\ref{fig:linelength}. Finally, 3000 $\mu$m long feedline is selected for subsequent simulation. 

\section{Calculation process of the impedance transformation networks}
\label{sec:ITN}

The calculation process of short step impedance transformation can be divided into four steps~\cite{1966Matthaei} .

Firstly, the ratio $r$ and the fractional bandwidth $w$ are calculated by the following formula:

\begin{equation}
\label{eq:2}
r=\frac{139.846\Omega}{50\Omega}=2.79692  ,  w=\frac{\theta_b - \theta_a}{\frac{\theta_b + \theta_a}{2}}=\frac{(8-4) GHz}{\frac{(8+4) GHz}{2}}=0.667
\end{equation}

Then the required $n$ value can be determined through the table of the pass band attenuation ripple $L_{Ar}$ in the article ~\cite{1966Matthaei} , and there are:

\begin{equation}
\label{eq:3}
n=10 , l=\frac{\lambda }{16}
\end{equation}

Furthermore, the impedance value of half of the circuit is directly obtained from the corresponding impedance table in the article~\cite{1966Matthaei} . And the residual impedance value are calculated by equation ~\ref{eq:4}.

\begin{equation}
\label{eq:4}
\begin{aligned}
Z_0=1 \ \ and \ \  Z_{n+1}=r
\\
Z_j|_{j=(n/2)+1\ to\ n}=\frac{r}{Z_{n+1-j}}
\end{aligned}
\end{equation}

All normalized resistance values are then converted to actual values. The feedline widths required by the impedance value are obtained through the calculation tool. Finally, the impedance transformation network with 10 segments of the same length but different widths is drawn.

% References
\bibliography{report} % bibliography data in report.bib

\begin{thebibliography}{10}

\bibitem{NAP26141}
{National Academies of Sciences, Engineering, and Medicine},  [{\em Pathways to
  Discovery in Astronomy and Astrophysics for the
  2020s}{\nolinebreak\hspace{0.1em}]}, The National Academies Press,
  Washington, DC (2021).

\bibitem{NASAexoplanets}
Brennan, P., ``Exoplanet catalog.'' [EB/OL] (2022).
\newblock \url{https://exoplanets.nasa.gov/discovery/exoplanet-catalog/}
  Accessed June 7, 2022.

\bibitem{benmazin19}
{Mazin}, B., {Bailey}, J., {Bartlett}, J., {Bockstiegel}, C., {Bumble}, B.,
  {Coiffard}, G., {Currie}, T., {Daal}, M., {Davis}, K., {Dodkins}, R.,
  {Fruitwala}, N., {Jovanovic}, N., {Lipartito}, I., {Lozi}, J., {Males}, J.,
  {Mawet}, D., {Meeker}, S., {O'Brien}, K., {Rich}, M., {Smith}, J., {Steiger},
  S., {Swimmer}, N., {Walter}, A., {Zobrist}, N., and {Zmuidzinas}, J.,
  ``{MKIDs in the 2020s},'' in [{\em Bulletin of the American Astronomical
  Society}{\nolinebreak\hspace{0.1em}]},   {\bf 51},  17 (Sept. 2019).

\bibitem{Mazin2013}
Mazin, B.~A., Meeker, S.~R., Strader, M.~J., Szypryt, P., Marsden, D., van
  Eyken, J.~C., Duggan, G.~E., Walter, A.~B., Ulbricht, G., Johnson, M.,
  Bumble, B., O'Brien, K., and Stoughton, C., ``{ARCONS}: A 2024 pixel optical
  through near-{IR} cryogenic imaging spectrophotometer,'' {\em Publications of
  the Astronomical Society of the Pacific}~{\bf 125},  1348--1361 (nov 2013).

\bibitem{Meeker2018}
Meeker, S.~R., Mazin, B.~A., Walter, A.~B., Strader, P., Fruitwala, N.,
  Bockstiegel, C., Szypryt, P., Ulbricht, G., Coiffard, G., Bumble, B.,
  Cancelo, G., Zmuda, T., Treptow, K., Wilcer, N., Collura, G., Dodkins, R.,
  Lipartito, I., Zobrist, N., Bottom, M., Shelton, J.~C., Mawet, D., van Eyken,
  J.~C., Vasisht, G., and Serabyn, E., ``{DARKNESS}: A microwave kinetic
  inductance detector integral field spectrograph for high-contrast
  astronomy,'' {\em Publications of the Astronomical Society of the
  Pacific}~{\bf 130},  065001 (apr 2018).

\bibitem{2019PhDW}
{Walter}, A.~B., {\em {MEC: The MKID Exoplanet Camera for High Speed Focal
  Plane Control at the Subaru Telescope}}, PhD thesis, University of
  California, Santa Barbara (Jan. 2019).

\bibitem{Walter2020}
Walter, A.~B., Fruitwala, N., Steiger, S., Bailey, J.~I., Zobrist, N., Swimmer,
  N., Lipartito, I., Smith, J.~P., Meeker, S.~R., Bockstiegel, C., Coiffard,
  G., Dodkins, R., Szypryt, P., Davis, K.~K., Daal, M., Bumble, B., Collura,
  G., Guyon, O., Lozi, J., Vievard, S., Jovanovic, N., Martinache, F., Currie,
  T., and Mazin, B.~A., ``The {MKID} exoplanet camera for subaru {SCExAO},''
  {\em Publications of the Astronomical Society of the Pacific}~{\bf 132},
  125005 (nov 2020).

\bibitem{Szypryt17}
{Szypryt}, P., {\em {Development of Microwave Kinetic Inductance Detectors for
  Applications in Optical to Near-IR Astronomy}}, PhD thesis, University of
  California, Santa Barbara (Jan. 2017).

\bibitem{2003Nature}
{Day}, P.~K., {LeDuc}, H.~G., {Mazin}, B.~A., {Vayonakis}, A., and
  {Zmuidzinas}, J., ``{A broadband superconducting detector suitable for use in
  large arrays},'' {\em Nature}~{\bf 425},  817--821 (Oct. 2003).

\bibitem{sonnet}
Rautio, J.~C., ``Sonnet software provides commercial eda software solutions for
  high-frequency rf/mw electromagnetic analysis..'' [EB/OL] (2022).
\newblock \url{https://www.sonnetsoftware.com/} Accessed June 7, 2022.

\bibitem{1966Matthaei}
{Matthaei}, G.~L., ``{Short-Step Chebyshev Impedance Transformers},'' {\em IEEE
  Transactions on Microwave Theory Techniques}~{\bf 14},  372--383 (Aug. 1966).

\bibitem{Jonas12}
Zmuidzinas, J., ``Superconducting microresonators: Physics and applications,''
  {\em Annual Review of Condensed Matter Physics}~{\bf 3},  169--214 (02 2012).

\bibitem{shu18}
Shu, S., Calvo, M., Leclercq, S., Goupy, J., Monfardini, A., and Driessen, E.,
  ``Prototype high angular resolution lekids for nika2,'' {\em Journal of Low
  Temperature Physics}~{\bf 193} (11 2018).

\bibitem{Shu2021}
Shu, S., Calvo, M., Goupy, J., Leclercq, S., Catalano, A., Bideaud, A.,
  Monfardini, A., and Driessen, E. F.~C., ``Understanding and minimizing
  resonance frequency deviations on a 4-in. kilo-pixel kinetic inductance
  detector array,'' {\em Applied Physics Letters}~{\bf 119},  092601 (aug
  2021).

\end{thebibliography}
\bibliographystyle{spiebib} % makes bibtex use spiebib.bst

\end{document}